\documentclass{article}
\usepackage[T1]{fontenc} 
\usepackage[utf8]{inputenc} 
\usepackage{ismir,amsmath,cite,url}
\usepackage{graphicx}
\usepackage{color}
\usepackage{amssymb}
\usepackage{amsthm}
\usepackage{xcolor}
\usepackage{enumitem}

\usepackage{fancyhdr}

\fancypagestyle{firstpage}{
  \fancyhf{} 
  \fancyhead[C]{\small An improved version of this paper has been published in TISMIR. See \href{https://transactions.ismir.net/articles/10.5334/tismir.251}{https://transactions.ismir.net/articles/10.5334/tismir.251}.}
}

\usepackage{lineno}

\title{PESTO: Pitch Estimation with Self-supervised Transposition-equivariant Objective}

\multauthor
{Alain Riou$^{1,2}$ \hspace{1cm} Stefan Lattner$^2$ \hspace{1cm} Gaëtan Hadjeres$^3$ \hspace{1cm} Geoffroy Peeters$^1$} {
	$^1$ LTCI, Télécom-Paris, Institut Polytechnique de Paris, France\\
	$^2$ Sony Computer Science Laboratories - Paris, France\\
	$^3$ Sony AI\\
	{\tt\small alain.riou@sony.com}
}

\def\authorname{A. Riou, S. Lattner, G. Hadjeres and G. Peeters}

\usepackage[bookmarks=false,pdfauthor={\authorname},pdfsubject={\papersubject},hidelinks]{hyperref}

\hypersetup{
    colorlinks=true,
    linkcolor=red,
    citecolor=green,
    filecolor=magenta,      
    urlcolor=blue,
    pdftitle={PESTO, Riou et al.},
    pdfpagemode=FullScreen,
    }

\graphicspath{{figures/}}

\newcommand{\LL}{\mathcal{L}}

\newcommand{\RR}{\mathbb{R}}

\newcommand{\TT}{\mathcal{T}}
\newcommand{\XX}{\mathcal{X}}
\newcommand{\YY}{\mathcal{Y}}

\newcommand{\CE}{\text{CrossEntropy}}

\newcommand{\f}{f_{\theta}}
\newcommand{\x}{\bold{x}}
\newcommand{\xk}{\bold{x}^{(k)}}
\newcommand{\txk}{\tilde{\bold{x}}^{(k)}}
\newcommand{\y}{\bold{y}}
\newcommand{\yk}{\bold{y}^{(k)}}
\newcommand{\tyk}{\tilde{\bold{y}}^{(k)}}

\newcommand{\Lequiv}{\mathcal{L}_{\text{equiv}}}
\newcommand{\Linv}{\mathcal{L}_{\text{inv}}}
\newcommand{\Lreg}{\mathcal{L}_{\text{SCE}}}

\newcommand{\xvoc}{\bold{x}_{\text{vocals}}}
\newcommand{\xback}{\bold{x}_{\text{background}}}

\usepackage{acronym}
\acrodef{mir}[MIR]{Music Information Retrieval}
\acrodef{cqt}[CQT]{Constant-Q Transform}
\acrodef{rpa}[RPA]{\emph{Raw Pitch Accuracy}}
\acrodef{rca}[RCA]{\emph{Raw Chroma Accuracy}}
\acrodef{ppa}[PPA]{\emph{Precise Pitch Accuracy}}
\acrodef{ta}[TA]{\emph{Tone Accuracy}}
\acrodef{ssl}[SSL]{Self Supervised Learning}

\DeclareMathOperator*{\argmax}{arg\,max}

\newtheorem*{definition}{Definition}

\begin{document}

\maketitle
\thispagestyle{firstpage}
\begin{abstract}
    In this paper, we address the problem of pitch estimation using \ac{ssl}. The \ac{ssl} paradigm we use is equivariance to pitch transposition, which enables our model to accurately perform pitch estimation on monophonic audio after being trained only on a small unlabeled dataset.
    We use a lightweight ($<$ 30k parameters) Siamese neural network that takes as inputs two different pitch-shifted versions of the same audio represented by its \acl{cqt}. To prevent the model from collapsing in an encoder-only setting, we propose a novel class-based transposition-equivariant objective which captures pitch information. Furthermore, we design the architecture of our network to be transposition-preserving by introducing learnable Toeplitz matrices.

    We evaluate our model for the two tasks of singing voice and musical instrument pitch estimation and show that our model is able to generalize across tasks and datasets while being lightweight, hence remaining compatible with low-resource devices and suitable for real-time applications.
    In particular, our results surpass self-supervised baselines and narrow the performance gap between self-supervised and supervised methods for pitch estimation.
\end{abstract}

\section{Introduction} 

Pitch estimation is a fundamental task in audio analysis, with numerous applications, e.g. in \ac{mir} and speech processing.
It involves estimating the fundamental frequency of a sound, which allows to estimate its perceived pitch. 
Over the years, various techniques have been developed for pitch estimation, ranging from classical methods (based on signal processing)~\cite{RAPT,PRAAT,pYIN,SWIPE} to machine learning approaches \cite{Lee2012,Han2014}.

In recent years, deep learning has emerged as a powerful tool for a wide range of applications, outperforming classical methods in many domains. 
This is notably true in \ac{mir}, where deep learning has led to significant advances in tasks such as music transcription \cite{Hawthorne2018,Kim2019,Kong2021}, genre classification \cite{Song2017,Oramas2018,Ndou2021}, and instrument recognition \cite{Lostanlen2016,Han2017,Solanki2022}.
Pitch estimation has also benefited greatly from deep learning techniques \cite{CREPE,Weiss2022}.
However, these deep learning models often require a large amount of labelled data to be trained
, and can be computationally expensive, hindering their practical applications in devices with limited computing power and memory capabilities.
Additionally, these models are often task-specific and may not generalize well to different datasets or tasks \cite{ShortcutLearning}. 
Therefore, there is a need for a lightweight and generic model that does not require labelled data to be trained. 
We address this here.

%
We take inspiration from the equivariant pitch estimation \cite{SPICE} and the equivariant tempo estimation \cite{Quinton2022} algorithms which we describe in part \ref{sec:related}.
As those, we use a \ac{ssl} paradigm based on Siamese networks and equivariance to pitch transpositions (comparing two versions of the same sound that have been transposed by a random but known pitch shift). 
We introduce a new equivariance loss that enforces the model to capture pitch information specifically.

This work has the following \textbf{contributions}:
\begin{itemize}[nosep]
\itemsep=+3pt
\item we formulate pitch estimation as a multi-class problem (part~\ref{sec:objective}); while \cite{SPICE,Quinton2022} model pitch/tempo estimation as a regression problem,
\item we propose a novel class-based equivariance loss (part~\ref{sec:objective}) which prevents collapse; while \cite{SPICE} necessitates a decoder,
\item  the architecture of our model is lightweight and transposition-equivariant by design. For this, we introduce Toeplitz fully-connected layers (part~\ref{sec:archi}).
\end{itemize}
We evaluate our method on several datasets and show that it outperforms self-supervised baselines on single pitch estimation (part~\ref{part_results}).
We demonstrate the robustness of our method to domain-shift and background music, highlighting its potential for real-world applications (part~\ref{part_background}).

Our proposed method requires minimal computation resources and is thus accessible to a wide range of users for both research and musical applications.
In consideration of accessibility and reproducibility, we make our code and pretrained models publicly available\footnote{\url{https://github.com/SonyCSLParis/pesto}}.

\section{Related works}\label{sec:related}

\subsection{SSL to learn invariant representations.}

\textbf{Siamese networks.} 
Most common techniques for \ac{ssl} representation involve Siamese networks~\cite{Siamese}. See section \ref{sec:related}.
The underlying idea is to generate two views of an input, feed them to a neural network, and train the network by applying a criterion between the output embeddings.
Various techniques have been developed for generating views\footnote{The most common technique involves randomly applying data augmentations to inputs to create pairs of inputs that share semantic content.}. 

\textbf{Collapse.}
However, a major issue with these methods is ``collapse'', when all inputs are mapped to the same embedding. 
To address this, various techniques have been proposed.
One of the most common is SimCLR~\cite{SimCLR} which also uses negative samples to ensure that embeddings are far apart through a contrastive loss. 
Additionally, several regularization techniques have been developed that minimize a loss over the whole batch. Barlow Twins \cite{BarlowTwins} force the cross-correlation between embeddings to be identity, while VICReg \cite{VICReg} add loss terms on the statistics of a batch to ensure that dimensions of the embeddings have high enough variance while remaining independent of each other. On the other hand, \cite{Wang2020a} explicitly minimize a loss over the hypersphere to distribute embeddings uniformly.
%
%
Furthermore, incorporating asymmetry between inputs has been shown to improve performance. \cite{MoCo, MoCov2} uses a momentum encoder, while \cite{BYOL} and \cite{SimSiam} add a projection head and a stop-gradient operator on top of the network, with \cite{BYOL} also using a teacher network. Finally, \cite{Dubois2022} incorporates asymmetry to contrastive- and clustering-based representation learning.

\textbf{Application to audio.}
While originally proposed for computer vision, these methods have been successfully adapted to audio and music as well. 
For example, \cite{COLA}, \cite{AudioBarlowTwins}, and \cite{BYOLA} respectively adapted \cite{SimCLR}, \cite{BarlowTwins}, and \cite{BYOL} to the audio domain. 
By training their large models on AudioSet~\cite{AudioSet}, they aim at learning general audio representations that are suited for many downstream tasks. More specifically, \cite{CLMR} successfully adapts contrastive learning to the task of music tagging by proposing more musically-relevant data augmentations.


\subsection{SSL to learn equivariant representations.}

The purpose of the methods described above is to learn a mapping $f : \XX \to \YY$ that is \textit{invariant} to a set of transforms $\TT_\XX$, i.e. so that for any input $\x \in \XX$ and transform $t \in \TT_\XX$
\begin{equation}
    f(t(\x)) \approx f(\x)
\end{equation}

However, recent approaches \cite{Dangovski2021,Winter2022,SIE} try instead to learn a mapping $f$ that is \emph{equivariant} to $\TT_\XX$, i.e. that satisfies
\begin{equation}
    f(t(\x)) \approx t'(f(\x))
\end{equation}
where $t' \in \TT_\YY$ with $\TT_\YY$ a set of transforms
that acts on the output space $\YY$. 
In other words, if the input is transformed, the output should be transformed accordingly. 
Representation collapse is hence prevented by design.

Equivariant representation learning has mostly been applied to computer vision and usually combines an invariance and an equivariance criterion.
E-SSL \cite{Dangovski2021} trains two projection heads on top of an encoder, one to return projections invariant to data augmentations while the other predicts the parameters of the applied data augmentations.
\cite{Winter2022} predicts separately a semantic representation and a rotation angle of a given input and optimizes the network with a reconstruction loss applied to the decoded content representation rotated by the predicted angle.
Finally, SIE \cite{SIE} creates a pair of inputs by augmenting an input and learns equivariant representations by training a hypernetwork conditioned on the parameters of the augmentation to predict one embedding of the pair from the other.

\textbf{Application to audio.}
Finally, a few successful examples of equivariant learning for solving MIR tasks recently emerged \cite{SPICE, Quinton2022}.
In particular, \cite{Quinton2022} introduces a simple yet effective equivariance criterion for tempo estimation while preventing collapse without any decoder or regularization: pairs are created by time-stretching an input with two different ratios, then the output embeddings are linearly projected onto scalars and the network is optimized to make the ratio of the scalar projections match the time-stretching ratio within a pair.

\subsection{Pitch estimation.}
Monophonic pitch estimation has been a subject of interest for over fifty years~\cite{Noll1967}. 
The earlier methods typically obtain a pitch curve by processing a candidate-generating function such as cepstrum~\cite{Noll1967}, autocorrelation function (ACF)~\cite{ACF}, and average magnitude difference function (AMDF)~\cite{AMDF}.
Other functions, such as the normalized cross-correlation function (NCCF)~\cite{PRAAT,RAPT} and the cumulative mean normalized difference function~\cite{YIN,pYIN}, have also been proposed.
On the other hand, \cite{SWIPE} performs pitch estimation by predicting the pitch of the sawtooth waveform whose spectrum best matches the one of the input signal.

Recently, methods involving machine learning techniques have been proposed~\cite{Lee2012,Han2014}. 
In particular, CREPE~\cite{CREPE} is a deep convolutional network trained on a large corpus to predict pitch from raw audio waveforms.
SPICE~\cite{SPICE} is a self-supervised method that takes as inputs individual \ac{cqt} frames of pitch-shifted inputs and learns the transposition between these inputs. 
It achieves quite decent results thanks to a decoder that takes as input the predicted pitch and tries to reconstruct the original \ac{cqt} frame from it.

Finally, some works~\cite{Luo2020,Tanaka2022} aim at disentangling the pitch and timbre of an input audio, thus predicting pitch as a side effect.
In particular, DDSP-inv~\cite{DDSPinv} is a DDSP-based approach \cite{DDSP} that relies on inverse synthesis to infer pitch in a self-supervised way.

\section{Self-supervised pitch estimation}

\subsection{Transposition-equivariant objective}\label{sec:objective}

We focus on the problem of monophonic pitch estimation and model it as a classification task. 
Our model is composed of a neural network $\f$ that takes as input an audio signal $\x$ and returns a vector $\y = (y_0, \dots, y_i, \dots, y_{d-1}) \in [0, 1]^d$, which represents the probability distribution of each pitch~$i$.
$y_i$ represents the probability that $i$ is the pitch of $\x$.
We propose here to train $\f$ in a \ac{ssl} way.
For this, similarly to \cite{SimCLR, BYOL, VICReg, MoCo, SimSiam}, we use data augmentations and Siamese networks.

Given $\x$, we first generate ${\x}^{(k)}$ by pitch-shifting $\x$ by a \emph{known} number $k$ of semitones.
Then, both $\x$ and ${\x}^{(k)}$ are fed to $\f$ which is trained to minimize a loss function between $\y\!=\!\f(\x)$ and ${\y}^{(k)}\!=\!\f({\x}^{(k)})$.

\begin{definition}
    For two vectors $\y, \y' \in \RR^d$ and $0 \leq k < d$, ${\y}'$ is a $k$-transposition of $\y$ if and only if for all $0 \leq i < d$
    \begin{equation}
        \begin{cases}
            y'_{i+k} = y_i	&\text{\emph{when} } \ 0 \leq i < d - k \\
            y'_i = 0		&\text{\emph{when} } \ i < k \\
            y_i = 0			&\text{\emph{when} } \ i \geq d - k - 1
        \end{cases}
    \end{equation}
    Similarly, for $-d < k \leq 0$, ${\y}'$ is a $k$-transposition of $\y$ if and only if $\y$ is a $-k$-transposition of ${\y}'$.
\end{definition}

\begin{figure}
    \centering
    \includegraphics[width=0.47\textwidth]{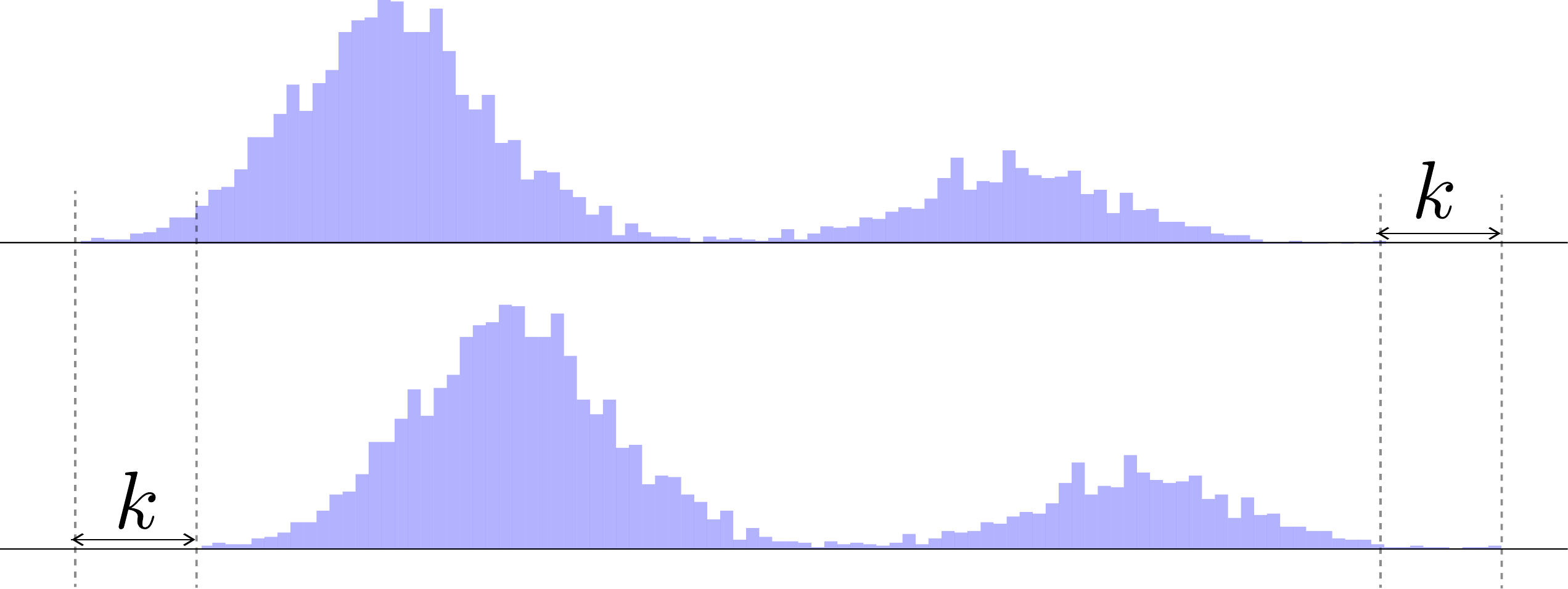}
    \caption{Example of $k$-transpositions. Visually, $\y$ and $\y'$ are just translated versions of each other. The sign of $k$ and its absolute value respectively indicate the direction and the distance of the translation.}
    \label{fig:ktransposition}
\end{figure}

The concept of $k$-transposition is illustrated in Figure~\ref{fig:ktransposition}.
Note also that for a vector $\y \in \RR^d$, exists at most one vector ${\y}' \in \RR^d$ that is a $k$-transposition of $\y$. 
We can therefore refer to ${\y}'$ as \emph{the} $k$-transposition of this vector $\y$.

\begin{figure*}
    \centering
    \includegraphics[width=\textwidth]{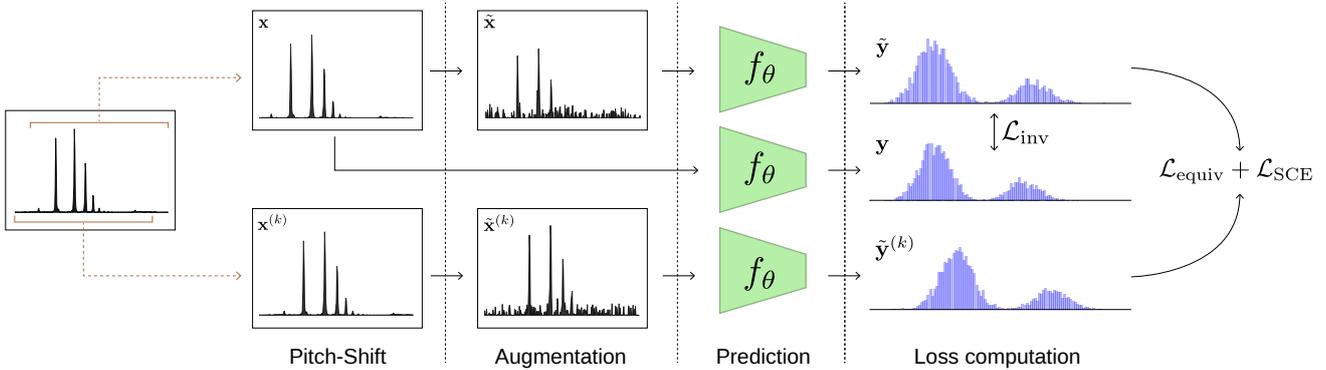}
    \caption{Overview of the PESTO method. The input \ac{cqt} frame (log-frequencies) is first cropped to produce a pair of pitch-shifted inputs $(\x, \xk)$. Then we compute $\tilde{\x}$ and $\txk$ by randomly applying pitch-preserving transforms to the pair. We finally pass $\x$, $\tilde{\x}$ and $\txk$ through the network $\f$ and optimize the loss between the predicted probability distributions.}
    \label{fig:model}
\end{figure*}

\textbf{Equivariance loss.}
We then design our criterion based on the following assumption: the probability of $\x$ to have pitch $i$ is equal to the probability of ${\x}^{(k)}$ to have pitch $i+k$, i.e. $y_i$ should be equal to $y_{i+k}^{(k)}$\footnote{For example, if $k = 2$ semitones, the probability of $\x$ to be C4 is exactly the probability of ${\x}^{(k)}$ to be a D4, and the same holds for any pitch independently of the actual pitch of $\x$.}. 
In other words, if ${\x}^{(k)}$ is a pitch-shifted version of $\x$, their respective pitch probability distributions should be shifted accordingly, i.e. ${\y}^{(k)}$ should be the $k$-transposition of $\y$.

We take inspiration from \cite{Quinton2022} to design our equivariance loss. However, in our case, the output of our network $\f$ is not a generic representation but a probability distribution.
We therefore adapt our criterion by replacing the learnable linear projection head from \cite{Quinton2022} by the following \emph{deterministic} linear form:
\begin{equation}
    \begin{matrix}
        \phi & : & \RR^d & \to & \RR \\
        &   & \y & \mapsto & (\alpha, \alpha^2, \dots, \alpha^d) \y
    \end{matrix}
\end{equation}
where $\alpha$ is a fixed hyperparameter\footnote{We found $\alpha = 2^{1/36}$ to work well in practice.}.

Indeed, with this formulation, for any $k$ if ${\y}'$ is a $k$-transposition of $\y$ then $\phi({\y}') = \alpha^k \phi(\y)$.
Hence we define our loss as
\begin{equation}
    \Lequiv(\y, {\y}^{(k)}, k) = h_{\tau}\left( \dfrac{\phi({\y}^{(k)})}{\phi(\y)} - \alpha^k \right)
\end{equation}
where $h_\tau$ is the Huber loss function \cite{Huber1964}, defined by
\begin{equation}
    h_{\tau}(x) =
    \begin{cases}
        \frac{x^2}{2} \ \text{ if } |x| \leq \tau \\
        \frac{\tau^2}{2} + \tau(|x| - \tau) \ \text{ otherwise}
    \end{cases}
\end{equation}

\textbf{Regularization loss.}
Note that if $\yk$ is the $k$-transposition of $\y$ then $\Lequiv(\y, \yk, k)$ is minimal. However, the converse is not always true.
In order to actually enforce pitch-shifted pairs of inputs to lead to $k$-transpositions, we further add a regularization term which is simply the shifted cross-entropy (SCE) between $\y$ and $\yk$, i.e. the cross-entropy between the $k$-transposition of $\y$ and $\yk$:
\begin{equation}
    \Lreg(\y, {\y}^{(k)}, k) = \sum_{i=0}^{d-1} y_i \log \left( y^{(k)}_{i+k} \right)
    \label{eq:entropy}
\end{equation}
with the out-of-bounds indices replaced by 0.
The respective contribution of $\Lequiv$ and $\Lreg$ is studied in part~\ref{sec:ablation}.

\textbf{Invariance loss.}
$\LL_{\text{equiv}}$ and $\Lreg$ allow our model to learn relative transpositions between different inputs and learn to output probability distributions $\y$ and ${\y}^{(k)}$ that satisfy the equivariance constraints.
However, these distributions may still depend on the timbre of the signal.
This is because our model actually never observed at the same time two different samples with the same pitch.

To circumvent this, we rely on a set $\TT$ of data augmentations that preserve pitch (such as gain or additive white noise).
We create augmented views $\tilde{\x} = t(\x)$ of our inputs $\x$ by applying random transforms $t \sim \TT$.

Similarly to \cite{CLMR}, we then train our model to be invariant to those transforms by minimizing the cross-entropy between $\y = \f(\x)$ and $\tilde{\y} = \f(\tilde{\x})$.
\begin{equation}
    \Linv(\y, \tilde{\y}) = \CE(\y, \tilde{\y})
\end{equation}

\textbf{Combining the losses.}
For a given input sample $\x$ and a given set of augmentations $\TT$, 
\begin{itemize}
\itemindent=-15pt
\itemsep=-3pt
\item we first compute $\xk$ by pitch-shifting $\x$ by a random number of bins $k$ (the precise procedure is described in section \ref{sec:frontend}); 
\item we then generate two augmented views $\tilde{\x} = t_1(\x)$ and $\txk = t_2(\xk)$, where $t_1, t_2 \sim \TT$;
\item we compute $\y\!=\!\f(\x)$, $\tilde{\y}\!=\!\f(\tilde{\x})$ and $\tyk\!=\!\f(\txk)$.
\end{itemize}

Our final objective loss is then:
\begin{equation}
    \begin{aligned}
        \LL(\y, \tilde{\y}, \tyk, k)
        &= \lambda_{\text{inv}} \;\;  \Linv(\y, \tilde{\y}) \\
        &+ \lambda_{\text{equiv}} \;\; \Lequiv(\tilde{\y}, \tyk, k) \\
        &+ \lambda_{\text{SCE}} \;\; \Lreg(\tilde{\y}, \tyk, k)
    \end{aligned}
\end{equation}

We illustrate this in Figure~\ref{fig:model}.
To set the weights $\lambda_{*}$ we use the gradient-based method proposed by \cite{GradNorm, VQGAN, MacGlashan2022}. 



\subsection{Audio-frontend}\label{sec:frontend}

The inputs $\x$ are the individual frames of the \ac{cqt}.
We have chosen the \ac{cqt} as input since its logarithmic frequency scale, in which bins of the \ac{cqt} exactly correspond to a fixed fraction $b$ of pitch semitones, naturally leads to pitch-shifting by translation.
\ac{cqt} is also a common choice made for pitch estimation \cite{SPICE, Weiss2022, BasicPitch}.



To compute the \ac{cqt}, we use the implementation provided in the nnAudio library \cite{nnAudio} since it supports parallel GPU computation. 
We choose $f_{\min} = 27.5$ Hz, which is the frequency of A0 the lowest key of the piano and select a resolution of $b = 3$ bins per semitone. 
Our \ac{cqt} has in total $K = 99b$ log-frequency bins, which corresponds to the maximal number of bins for a 16kHz signal. 

\subsection{Simulating translations.}\label{sec:simulating}

To avoid any boundary effects, we perform pitch-shift by cropping shifted slices of the original \ac{cqt} input frame as in \cite{SPICE}\footnote{Specifically, we sample an integer $k$ uniformly from the range $\{-k_{max}, \dots, k_{\max}\}$, then generate two CQT outputs, denoted as $x$ and $x^{(k)}$, where $x$ is obtained by cropping the input CQT at indices $[k_{\max}, K-k_{\max}-1]$, and $x^{(k)}$ is obtained by cropping the input CQT at indices $[k_{\max}-k, K-k_{\max}+k-1]$, with $K$ the total number of bins of the original CQT frame and $k_{\max}$ = 16 in practice (see Figure \ref{fig:model}).}.
From a computational point of view, it is indeed significantly faster than applying classical pitch shift algorithms based on phase vocoder and resampling.

\subsection{Transpostion-preserving architecture}\label{sec:archi}

\begin{figure}
    \centering
    \includegraphics[width=0.48\textwidth]{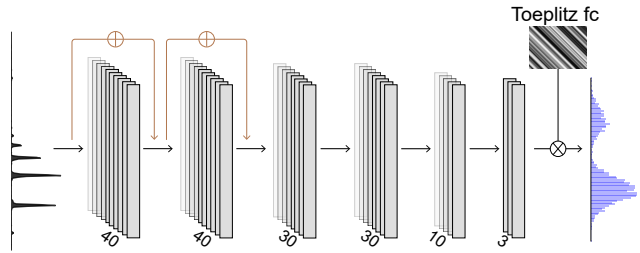}
    \caption{Architecture of our network $\f$. The number of channels varies between the intermediate layers, however the frequency resolution remains unchanged until the final Toeplitz fully-connected layer.}
    \label{fig:archi}
\end{figure}

The architecture of $\f$ is illustrated in Figure~\ref{fig:archi}. It is inspired by~\cite{Weiss2022}.
Each input \ac{cqt} frame is processed independently: first layer-normed \cite{LayerNorm} then preprocessed by two 1D-Conv (convolution in the log-frequency dimension) with skip-connections~\cite{ResNet}, followed by four 1D-Conv layers.
As in \cite{Weiss2022}, we apply a non-linear leaky-ReLU (slope 0.3)~\cite{LeakyReLU} and dropout (rate 0.2)~\cite{Dropout} between each convolutional layer. 
Importantly, the kernel size and padding of each of these layers are chosen so that the frequency resolution is never reduced.
We found in practice that it helps the model to distinguish close but different pitches.
The output is then flattened, fed to a final fully-connected layer and normalized by a softmax layer to become a probability distribution of the desired shape.

Note that all layers (convolutions\footnote{Convolutions roughly preserve transpositions since the kernels are applied locally, meaning that if two transposed inputs are convolved by the same kernel, then the output results will be almost transpositions of each other as well.}, elementwise non-linearities, layer-norm and softmax), except the last final fully-connected layer, preserve transpositions.
To make the final fully-connected layer also transposition-equivariant, we propose to use \textbf{Toeplitz fully-connected layers}. It simply consists of a standard linear layer without bias but whose weights matrix $A$ is a Toeplitz matrix, i.e. each of its diagonals is constant.
    \begin{equation}
    \label{eq_toeplitz}
    A =
    \begin{pmatrix}
        a_0	    & a_{-1} & a_{-2} & \cdots & a_{-n+2} & a_{-n+1} \\
        a_1	    & a_0    & a_{-1} & \ddots & \ddots & a_{-n+2} \\
        a_2     & a_1    & \ddots & \ddots & \ddots & \vdots \\
        \vdots  & \ddots & \ddots & \ddots & \ddots	& \vdots \\
        a_{m-1} & \cdots & \cdots & \cdots & \cdots	& a_{m-n}
    \end{pmatrix}
\end{equation}
    Contrary to arbitrary fully-connected layers, Toeplitz matrices are transposition-preserving operations and only have $m + n - 1$ parameters instead of $mn$.
    Furthermore, they are mathematically equivalent to convolutions, making them straightforward to implement.



\begin{table*}[t]
    \centering
    \begin{tabular}{lcccc}
        \hline
        &  &  & \multicolumn{2}{c}{Raw Pitch Accuracy} \\
        Model					& \# params		& Trained on			& \emph{MIR-1K}	& \emph{MDB-stem-synth}	\\
        \hline
        SPICE \cite{SPICE}  	& 2.38M			& private data		& 90.6\%		   & 89.1\%		\\
        DDSP-inv \cite{DDSPinv} & 21.6M	& \emph{MIR-1K} / \emph{MDB-stem-synth}			& 91.8\%		& 88.5\%			\\
        \hline
        PESTO (ours) &
        28.9k &
        \emph{MIR-1K} &
        \textbf{96.1\%} &
        94.6\% \\
        
        PESTO (ours) &
        28.9k &
        \emph{MDB-stem-synth} &
        93.5\% &
        \textbf{95.5\%} \\
        \hline
        \color{gray}CREPE \cite{CREPE} &
        \color{gray}22.2M &
        \color{gray}many (supervised) &
        \color{gray}\textbf{97.8\%} &
        \color{gray}\textbf{96.7\%} \\
        \hline
    \end{tabular}
    \caption{Evaluation results of PESTO compared to supervised and self-supervised baselines. CREPE has been trained in a supervised way on a huge dataset containing in particular \emph{MIR-1K} and \emph{MDB-stem-synth}. It is grayed out as a reference. For DDSP-inv, we report the results when training and evaluating on the same dataset.}
    \label{tab:results}
\end{table*}

\subsection{Absolute pitch inference from $\y$}\label{sec:absolute}

Our encoder $\f$ returns a probability distribution over (quantized) pitches. 
From an input \ac{cqt} frame $\x$, we first compute the probability distribution $\f(\x)$, then we infer the absolute pitch $\hat{p}$ by applying the affine mapping:
\begin{equation}
    \hat{p}(\x) = \dfrac{1}{b} \left( \argmax \f(\x) + p_0 \right)
\end{equation}
where $b = 3$ is the number of bins per semitones in the \ac{cqt} and $p_0$ is a fixed integer shift that only depends on $\f$.
As in \cite{SPICE}, we set the integer shift $p_0$ by relying on a set of synthetic data\footnote{synthetic harmonic signals with random amplitudes and pitch} with known pitch.






\section{Experiments}

\subsection{Datasets}

To evaluate the performance of our approach, we consider the two following datasets:
\begin{enumerate}[nosep]
    \item \textit{MIR-1K} \cite{MIR-1K} contains 1000 tracks (about two hours) of people singing Chinese pop songs, with separate vocal and background music tracks provided.
    \item \textit{MDB-stem-synth} \cite{MDBstemsynth} contains re-synthesized monophonic music played by various instruments.
\end{enumerate}

The pitch range of the \textit{MDB-stem-synth} dataset is wider than the one of \textit{MIR-1K}. 
The two datasets have different sampling rates and granularity for the annotations.

We conduct separate model training and evaluation on both datasets to measure overfitting and generalization performance.
In fact, given that our model is lightweight and does not require labelled data, overfitting performance is particularly relevant for real-world scenarios, as it is easy for someone to train on their own dataset, e.g. their own voice.
However, we also examine generalization performance through cross-evaluation to ensure that the model truly captures the underlying concept of pitch and does not merely memorize the training data.

\subsection{Training details}

From an input \ac{cqt} (see part~\ref{sec:frontend}), we first compute the pitch-shifted \ac{cqt} (see part~\ref{sec:simulating}).
Then two random data augmentations $t_1, t_2 \sim \TT$ are applied with a probability of 0.7.
We used white noise with a random standard deviation between 0.1 and 2, and gain with a random value picked uniformly between -6 and 3 dB.
The overall architecture of $\f$ (see part~\ref{sec:archi}) is implemented in PyTorch~\cite{PyTorch}.
For training, we use a batch size of 256 and the Adam optimizer \cite{Adam} with a learning rate of $10^{-4}$ and default parameters. The model is trained for 50 epochs using a cosine annealing learning rate scheduler.
Our architecture being extremely lightweight, training requires only 545MB of GPU memory and can be performed on a single GTX 1080Ti.

\subsection{Performance metrics}

We measure the performances using the following metrics.
\begin{enumerate}[nosep]
    \item \ac{rpa}: corresponds to the percentage of voiced frames whose pitch error\footnote{i.e. distance between the predicted pitch and the actual one} is less than 0.5 semitone~\cite{Poliner2007}.
    \item \ac{rca}: same as \ac{rpa} but considering the mapping to Chroma (hence allowing octave errors)~\cite{Poliner2007}.
\end{enumerate}

\ac{rca} is only used in our ablation studies. 

\subsection{Results and discussions}
\label{part_results}

\subsubsection{Clean signals}
\label{part_results}

We compare our results with three baselines: CREPE~\cite{CREPE}, SPICE~\cite{SPICE} and DDSP-inv~\cite{DDSPinv}.
CREPE is fully-supervised while SPICE and DDSP-inv are two \ac{ssl} approaches.
To measure the influence of the training set, we train PESTO on the two datasets (\textit{MIR-1K} and \textit{MDB-stem-synth}) and also evaluate on the two. 
This allows to test model generalization.

We indicate the results in Table~\ref{tab:results}.
We see that PESTO significantly outperforms the two \ac{ssl} baselines (SPICE and DDSP-inv) even in the cross-dataset scenario (93.5\% and 94.6\%).
Moreover, it is competitive with CREPE (-1.7\% and -1.2\%) which has 750 times more parameters and is trained in a supervised way on the same datasets.


\subsubsection{Robustness to background music}
\label{part_background}

\begin{table}[]
    \centering
    \begin{tabular}{lcccc}
        \hline
        & \multicolumn{4}{c}{Raw Pitch Accuracy (\textit{MIR-1K})} \\
        Model	& clean				& 20 dB				& 10 dB				& 0 dB				\\
        \hline
        SPICE\cite{SPICE} &
        91.4\% &
        91.2\% & $90.0\%$			& $81.6\%$			\\
        \hline
        
        PESTO	& & & \\
        $\beta = 0$ &
        \textbf{94.8\%} &
        90.7\% &
        79.2\% &
        50.0\% \\
        
        $\beta = 1$ &
        94.5\% &
        94.2\% &
        92.9\% &
        \textbf{83.1\%} \\
        
        $\beta \sim \mathcal{U}(0,1)$ &
        94.7\% &
        94.4\% &
        92.9\% &
        81.7\% \\
        
        $\beta \sim \mathcal{N}(0, 1)$ &
        \textbf{94.8\%} &
        \textbf{94.5\%} &
        \textbf{93.0\%} &
        82.6\% \\
        
        $\beta \sim \mathcal{N}(0, \frac{1}{2})$ &
        \textbf{94.8\%} &
        \textbf{94.5\%} &
        92.9\% &
        81.0\% \\
        \hline
        
        \color{gray}CREPE\cite{CREPE} &
        \color{gray}\textbf{97.8\%} &
        \color{gray}\textbf{97.3\%} &
        \color{gray}\textbf{95.3\%} &
        \color{gray}\textbf{84.8\%} \\
        \hline
    \end{tabular}
    \caption{Robustness of PESTO and other baselines to background music with various Signal-to-Noise ratios. Adding background music to training samples significantly improves the robustness of PESTO (see section \ref{part_background}).}
    \label{tab:background}
\end{table}

Background noise and music can severely impact pitch estimation algorithms, making it imperative to develop robust methods that can handle real-world scenarios where background noise is often unavoidable.

We therefore test the robustness of PESTO to background music.
For this, we use the \textit{MIR-1K} dataset, which contains separated vocals and background tracks and allows testing various signal-to-noise (here vocal-to-background) ratios (SNRs). 

We indicate the results in Table~\ref{tab:background}.
As foreseen, the performance of PESTO when trained on clean vocals (row $\beta = 0$) and applied to vocal-with-background considerably drop: from 94.8\% (clean) to 50.0\% (SNR = 0 dB)\footnote{It should be noted that the difference between the 96.1\% of Table~\ref{tab:results} and the 94.8\% of Table~\ref{tab:background} is due to the fact that we do not apply any data augmentation (gain or additive white noise) when $\beta=0$.}.

To improve the robustness to background music, we slightly modify our method to train our model on mixed sources. 
Instead of using gain and white noise as data augmentations, we create an augmented view of our original vocals signal $\xvoc$ by mixing it (in the complex-CQT domain) with its corresponding background track $\xback$:
\begin{equation}
    \x = \xvoc + \beta \xback
\end{equation}
Then, thanks to $\Linv$, the model is trained to ignore the background music for making its predictions.

The background level $\beta$ is randomly sampled for each CQT frame. The influence of the distribution we sample $\beta$ from is depicted in Table \ref{tab:background}.
This method significantly limits the drop in performances observed previously and also makes PESTO outperform SPICE in noisy conditions.

\subsubsection{Ablation study}\label{sec:ablation}
~
Table~\ref{tab:ablation} depicts the influence of our different design choices. 
First, we observe that the equivariance loss $\Lequiv$ and the final Toeplitz fully-connected layer (Eq.~\ref{eq_toeplitz}) are absolutely essential for our model not to collapse. 
Moreover, data augmentations seem to have a negligible influence on out-of-domain RPA (-0.2\%) but slightly help when training and evaluating on the same dataset (+1.2\%).

On the other hand, it appears that both $\Linv$ and $\Lreg$ do not improve in-domain performances but help the model to generalize better. This is especially true for $\Lreg$, whose addition enables to improve RPA from 86.9\% to 94.6\% on \textit{MDB-stem-synth}.

Finally, according to the drop of performances in RPA and RCA when removing $\Linv$, it seems that the invariance loss prevents octave errors on the out-of-domain dataset.

\begin{table}
    \centering
        \begin{tabular}{lcccc}
        \hline
                                    & \multicolumn{2}{c}{MIR-1K}    & \multicolumn{2}{c}{MDB} \\
                                    & RPA		& RCA		& RPA		& RCA		\\
        \hline
        PESTO baseline				& 96.1\%	& 96.4\%	& 94.6\%	& 95.0\%	\\
        \hline
        \emph{Loss ablations}		&			&			&			&			\\
        w/o $\Lequiv$		& 5.8\%		& 8.6\%		& 1.3\%		& 6.1\%		\\
        w/o $\Linv$		& 96.1\%	& 96.4\%	& 92.5\%	& 94.5\%	\\
        w/o $\Lreg$	& 96.1\%	& 96.5\%	& 86.9\%	& 93.8\%	\\
        \hline
        \emph{Miscellaneous}		& \\
        no augmentations  			& 94.8\%    & 95.4\%    & 94.8\%    & 95.2\%    \\
        non-Toeplitz fc				& 5.7\%		& 8.7\%		& 1.2\%		& 6.1\%		\\
        \hline
    \end{tabular}
    \caption{Respective contribution of various design choices of PESTO for a model trained on \textit{MIR-1K}.}
    \label{tab:ablation}
\end{table}

\section{Conclusion}

In this paper, we presented a novel self-supervised learning method for pitch estimation that leverages equivariance to musical transpositions.
We propose a class-based equivariant objective that enables Siamese networks to capture pitch information from pairs of transposed inputs accurately. We also introduce a Toeplitz fully-connected layer to the architecture of our model to facilitate the optimization of this objective. Our method is evaluated on two standard benchmarks, and the results show that it outperforms self-supervised baselines and is robust to background music and domain shift.

From a musical perspective, our lightweight model is well-suited for real-world scenarios, as it can run on resource-limited devices without sacrificing performance. Moreover, its \ac{ssl} training procedure makes it convenient to fine-tune on a small unlabeled dataset, such as a specific voice or instrument.
Additionally, the resolution of the model is a sixth of a tone but could eventually be increased by changing the resolution of the CQT. Moreover, despite modelling pitch estimation as a classification problem, we make no assumption about scale or temperament.

These features make our method still a viable solution, e.g. for instruments that use quartertones and/or for which no annotated dataset exists. We therefore believe that it has many applications even beyond the limitations of Western music.\\

Overall, the idea of using equivariance to solve a classification problem is a novel and promising approach that enables the direct return of a probability distribution over the classes with a single, potentially synthetic, labelled element.
While our paper applies this approach to pitch estimation, there are other applications where this technique could be useful, such as tempo estimation.

Moreover, modelling a regression task as a classification problem 
can offer greater interpretability as the output of the network is not a single scalar but a whole probability distribution. Finally, it can generalize better to multi-label scenarios.

Our proposed method hence demonstrates the potential of using equivariance to solve problems that are beyond the scope of our current work. In particular, it paves the way towards self-supervised multi-pitch estimation.

\section{Acknowledgements}

This work has been funded by the ANRT CIFRE convention n°2021/1537 and Sony France. This work was granted access to the HPC/AI resources of IDRIS under the allocation 2022-AD011013842 made by GENCI. We would like to thank the reviewers and meta-reviewer for their valuable and insightful comments.

\bibliography{library}

%
%
%
%
%

\end{document}